\documentclass[letterpaper,12pt]{article}
\usepackage{array}
\usepackage{dcolumn}
\usepackage{float}
\usepackage{natbib}
\usepackage{graphicx}
\usepackage{color}
\usepackage{amsmath}
\usepackage{amsfonts}
\usepackage{amsopn}
\usepackage{amsthm}
\usepackage{amssymb}
\usepackage{lscape}
\usepackage{multirow}
\newtheorem{thm}{Theorem}
\newtheorem{lem}{Lemma}

\theoremstyle{definition}

\newtheorem{asmp}{Assumption}
\theoremstyle{remark}
\newtheorem{rmk}{Remark}
\newtheorem*{rmk.nonumber}{Remark}
\DeclareMathOperator{\pr}{pr}
\DeclareMathOperator{\epn}{E}
\DeclareMathOperator{\var}{var}

\DeclareMathOperator{\logit}{logit}

\newcommand{\bmW}{W}

\begin{document}
\begin{center}
{\LARGE Optimal Treatment Allocations Accounting for Population Differences}
\end{center}
\begin{center}
Wei Zhang$^{1}$, Zhiwei Zhang$^{2, *}$, and Aiyi Liu$^{3}$\\
$^1$State Key Laboratory of Mathematical Sciences, Academy of Mathematics and Systems Science, Chinese Academy of Sciences, Beijing, China\\
$^2$Biostatistics Innovation Group, Gilead Sciences, Foster City, California, USA\\
$^3$Biostatistics and Bioinformatics Branch, Division of Population Health Research, \textit{Eunice Kennedy Shriver} National Institute of Child Health and Human Development, National Institutes of Health, Bethesda, Maryland, USA\\
$^*$zhiwei.zhang6@gilead.com
\end{center}

\vspace{0.5cm}
\centerline{\bf Abstract}

The treatment allocation mechanism in a randomized clinical trial can be optimized by maximizing the nonparametric efficiency bound for a specific measure of treatment effect. Optimal treatment allocations which may or may not depend on baseline covariates have been derived for a variety of effect measures focusing on the trial population, the patient population represented by the trial participants. Frequently, clinical trial data are used to estimate treatment effects in a target population that is related to but different from the trial population. This article provides optimal treatment allocations that account for the impact of such population differences. We consider three cases with different data configurations: transportation, generalization, and post-stratification. Our results indicate that, for general effect measures, optimal treatment allocations may depend on the covariate distribution in the target population but not on the configuration of data or information that describes the target covariate distribution. For estimating average treatment effects, there is a unique covariate-dependent allocation that achieves maximal efficiency regardless of the target covariate distribution and the associated data configuration.

\vspace{.5cm}

\section{Introduction}\label{intro}

The treatment allocation mechanism in a randomized clinical trial can be optimized for statistical efficiency. A well-known result in this domain is the Neyman allocation \citep{n34}, which makes the size of each treatment group proportional to the standard deviation of the outcome of interest in the same treatment group. The Neyman allocation is optimal, in the sense of minimum asymptotic variance, for estimating the average treatment effect (ATE) with the sample mean difference. The sample mean difference is commonly used but not fully efficient in the presence of baseline covariates associated with treatment outcomes. Motivated by semiparametric theory \citep{b93,t06}, more efficient estimators have been developed to incorporate information from baseline covariates \citep[e.g.,][]{t08,z08,m09,r10,t12,zm19,y23}. Some of these estimators leverage modern machine learning and ensemble learning methods \citep[e.g.,][]{h09,p11} and have a realistic chance to attain or approach the nonparametric efficiency bound for treatment effect estimation. Optimal treatment allocations that maximize the nonparametric efficiency bound have been derived by \citet{z23}, who consider both the traditional covariate-independent randomization (CIR) design and a covariate-dependent randomization (CDR) design that resembles observational studies except that the propensity score is specified and known by the investigator. These efforts to improve statistical efficiency generally aim at a treatment effect measure, such as the ATE, in the trial population (i.e., the patient population represented by the trial participants).

Clinical trial data are frequently used to estimate treatment effects in a target patient population related to but different from the trial population. This is done for various reasons, and we give two examples here. First, the trial cohort is often a convenience sample and may not be truly representative of the patient population for treatment indication. If the latter population is better described by another data source, say a disease registry, it is of interest to combine trial data with registry information, which may be individual patient data or summary statistics, to estimate the effect of the experimental treatment in the patient population represented by the disease registry \citep{cs10,s11,d19,c21}. Second, calibrating the treatment effect observed in one trial to another trial population can be useful for making an indirect comparison of two treatments that have not been compared in a randomized fashion. For example, when an experimental treatment is evaluated against an active control treatment in a randomized non-inferiority trial, investigators and regulators often wonder how the experimental treatment would compare with a placebo control in the same setting. If the active control has been compared with placebo in a previous randomized trial, this question can be answered by calibrating the active control effect from the previous trial to the non-inferiority trial population \citep{z09,n13,z16}.

This article provides optimal treatment allocations for estimating treatment effects in a target population that overlaps with but differs from the trial population in terms of baseline covariates. We consider three cases with different data configurations. In the first case (transportation), individual covariate data are available in a target cohort (a random sample from the target population) which does not include any trial participants \citep[e.g.,][]{z16,rv17,c21}. The second case (generalization) is similar to the first one except that the target cohort contains the trial cohort as a subset \citep{s11,d19}. In the third case (post-stratification), the target population is defined by a set of known weights assigned to a finite number of strata such as demographic subgroups \citep{bd87,cs10}. In each case, we derive the optimal CIR and CDR designs, the latter being generally more efficient because CDR includes CIR as a special case. Interestingly, the optimal CIR and CDR designs can be characterized using general expressions that simultaneously apply to all three cases, indicating that optimal treatment allocations may depend on the covariate distribution in the target population but not on the configuration of data or information that describes the target covariate distribution. For estimating the ATE in the target population, the optimal CDR design does not even depend on the target covariate distribution and is identical to the optimal CDR design for estimating the ATE in the trial population \citep{z23}. Thus, for ATE estimation, the optimal CDR design has a desirable invariance property not shared by optimal CIR designs, in addition to an efficiency advantage.

The rest of the article is organized as follows. In the next section, we review existing results and present new ones on optimal treatment allocations. In Section \ref{nr}, we report numerical results from a simulation study and concerning a real example. Concluding remarks are given in Section \ref{cr}, and technical proofs provided in Supplementary Materials.

\section{Optimal Treatment Allocations}\label{ota}

\subsection{Preliminaries}\label{pre}

Suppose a randomized trial is to be conducted to compare an experimental treatment with a control treatment. For a generic patient in the trial, let $Y(a)$ denote the potential outcome for the experiment ($a=1$) or control ($a=0$) treatment, $W$ a vector of baseline covariates that may be associated with one or both potential outcomes, $A\in\{0,1\}$ the randomly assigned treatment, and $Y=Y(A)=AY(1)+(1-A)Y(0)$ the actual observed outcome. For treatment allocation, we consider both CIR and CDR, defined as
$$
\pr\{A=1|W,Y(0),Y(1)\}=\begin{cases}\pr(A=1)=\pi\in(0,1)&\text{(CIR)}\\
	\pr(A=1|W)=p(W)\in(0,1)&\text{(CDR)}\end{cases},
$$
where $p(W)$ is known as the propensity score for treatment assignment. CDR may look like an observational study but is fundamentally a randomized trial in that the treatment assignment mechanism is entirely controlled by the investigator. From a causal inference perspective, CIR and CDR provide the same level of statistical rigor as they both allow causal estimands to be identified and estimated without making strong assumptions \citep{z23}. The trial cohort is regarded as a random sample from some patient population, which we call the trial population. Accordingly, the observed trial data are conceptualized as independent copies of $(W,A,Y)$ and denoted by $(W_i,A_i,Y_i)$ $(i=1,\dots,n)$.

Our general objective is to optimize treatment allocation, as determined by $\pi$ or $p$, by maximizing the nonparametric efficiency bound for estimating a specified effect measure. Consider, for example, $\Delta=\epn\{Y(1)-Y(0)\}$, the ATE in the trial population. According to \citet{z23}, for estimating $\Delta$ using the trial data, the optimal CIR design is 
$$
\pi_{\textup{opt}}=\frac{[\epn\{v_1(W)\}]^{1/2}}{[\epn\{v_1(W)\}]^{1/2}+[\epn\{v_0(W)\}]^{1/2}},
$$
and the optimal CDR design by
\begin{equation}\label{opt.cdr}
	p_{\textup{opt}}(W)=\frac{v_1(W)^{1/2}}{v_1(W)^{1/2}+v_0(W)^{1/2}},
\end{equation}
where $v_a(W)=\var\{Y(a)|W\}$, $a=0,1$. The optimal CDR design can be seen as applying the Neyman allocation to each sub-population defined by $W$.

This article is mainly concerned with effect measures of the form $\Delta^*=\int\delta(w)\/\text{d}F^*(w)$, where $\delta(w)=m(1,w)-m(0,w)$, $m(a,w)=\epn\{Y(a)|W=w\}$, and $F^*$ is the covariate distribution in some target population. (Additional effect measures are considered briefly in Section \ref{ext}.) Under a certain exchangeability assumption \citep[e.g.,][]{z16,d19,c24}, $\Delta^*$ is the ATE in the target population. More generally, $\Delta^*$ can be interpreted as a calibrated or adjusted treatment effect that accounts for population differences. Let $F$ denote the distribution of $W$ in the trial population. Throughout, we make the following assumption:
\begin{asmp}\label{eq.supp}
	$F$ and $F^*$ dominate each other.
\end{asmp}
\noindent Let $r$ denote the density of $F^*$ with respect to $F$, which is also known as the density ratio of $F^*$ to $F$ with respect to a common dominating measure. Assumption \eqref{eq.supp} implies that $r$ exists and that $\pr\{0<r(W)<\infty\}=1$.  Assumption \ref{eq.supp} ensures that $\Delta^*$ is identifiable if $F^*$ is known or identifiable. Information about $F^*$ may come in different forms, as indicated in Section \ref{intro}. In the next three sections, we elaborate on the three cases mentioned in Section \ref{intro}. Any new notations we use to describe a case are only applicable within that context.

\begin{rmk}
	Assumption \ref{eq.supp} essentially states that $\mathcal W=\mathcal W^*$, where $\mathcal W$ (rsp. $\mathcal W^*$) denotes the support of $F$ (rsp. $F^*$). Technically, we only need $\mathcal W\supseteq\mathcal W^*$ for identifiability. If $\mathcal W\supsetneq\mathcal W^*$, trial participants with $W_i\in\mathcal W\setminus\mathcal W^*$ are not informative of $\Delta^*$ without making strong assumptions, and we therefore exclude such patients from consideration, effectively restricting $F$ to $\mathcal W^*$. As an example, consider estimating the ATE in a subpopulation $\mathcal W^*\subset\mathcal W$. For this purpose, the optimal value of $\pi$ is easily seen to be
	$$
	\frac{[\epn\{v_1(W)|W\in\mathcal W^*\}]^{1/2}}{[\epn\{v_1(W)|W\in\mathcal W^*\}]^{1/2}+[\epn\{v_0(W)|W\in\mathcal W^*\}]^{1/2}},
	$$
	and the optimal propensity score is  \eqref{opt.cdr} for $W\in\mathcal W^*$. In the event $\mathcal W\nsupseteq\mathcal W^*$, we restrict both distributions to the common support $\mathcal W\cap\mathcal W^*$, and redefine $(F,F^*)$ as the restricted distributions so as to satisfy Assumption \ref{eq.supp}.
\end{rmk}

\subsection{The Case of Transportation}\label{tr}

In this case, $F^*$ is described by observed covariate values (denoted by $W_j^*$, $j=1,\dots,n^*$) in a target cohort, a random sample from the target population that shares no patients with the trial cohort. We assume that the $W_j^*$'s are independent of each other and of $\{(W_i,A_i,Y_i),i=1,\dots,n\}$, and identically distributed according to $F^*$. For a generic patient in the trial or target cohort, let $Z$ be a source indicator ($Z=1$ if in the trial; 0 otherwise), and let $X=(W,A,Y)$ (if $Z=1$) or $W^*$ (if $Z=0$). Then the combined data, $\{(W_i,A_i,Y_i),i=1,\dots,n\}\cup\{W_j^*,j=1,\dots,n^*\}$, can be considered independent copies of $O=(Z,X)$.

Under CIR, \citet{z16} derived the efficient influence function for estimating $\Delta^*$ as
\begin{equation*}\begin{aligned}
		\psi^{\textup{tr}}_{\text{\sc cir}}(O)=\frac{(1-Z)\{\delta(W^*)-\Delta^*\}}{1-\gamma}
		+\frac{ZAr(W)\{Y-m(1,W)\}}{\gamma\pi}-\frac{Z(1-A)r(W)\{Y-m(0,W)\}}{\gamma(1-\pi)},
\end{aligned}\end{equation*}
where $\gamma=\pr(Z=1)$ and the superscript tr represents transportation. This result has motivated doubly robust, locally efficient estimators of $\Delta^*$ based on working parametric models for $(m,r)$ \citep{z16}. The nuisance functions $(m,r)$ can also be estimated using nonparametric machine learning methods, leading to nonparametric estimators of $\Delta^*$ that are consistent, asymptotically linear, and efficient under mild conditions \citep{c21}.

\begin{lem}
	In the case of transportation, the efficient influence function for estimating $\Delta^*$ under CDR is 
	\begin{equation*}\begin{aligned}
			\psi^{\textup{tr}}_{\text{\sc cdr}}(O)=\frac{(1-Z)\{\delta(W^*)-\Delta^*\}}{1-\gamma}
			+\frac{ZAr(W)\{Y-m(1,W)\}}{\gamma p(W)}-\frac{Z(1-A)r(W)\{Y-m(0,W)\}}{\gamma\{1-p(W)\}}.
	\end{aligned}\end{equation*}
\end{lem}

Thus, $\psi^{\textup{tr}}_{\text{\sc cdr}}(O)$ includes $\psi^{\textup{tr}}_{\text{\sc cir}}(O)$ as a special case with $p(W)\equiv\pi$. The existing estimators of $\Delta^*$, developed for CIR, can be extended to CDR while maintaining their essential properties. These extensions are straightforward and the details are omitted. 

The optimal CIR and CDR designs for transportantion can be derived by minimizing $\var\{\psi^{\textup{tr}}_{\text{\sc cir}}(O)\}$ and $\var\{\psi^{\textup{tr}}_{\text{\sc cdr}}(O)\}$ with respect to $\pi$ and $p$, respectively.

\begin{thm}
	Under Assumption \ref{eq.supp}, $\var\{\psi^{\textup{tr}}_{\text{\sc cir}}(O)\}$ is minimized uniquely by setting $\pi$ equal to
	\begin{equation}\label{opt.cir.tr}
		\pi_{\textup{opt}}^*=\frac{[\epn\{r(W)^2v_1(W)\}]^{1/2}}{[\epn\{r(W)^2v_1(W)\}]^{1/2}+[\epn\{r(W)^2v_0(W)\}]^{1/2}},
	\end{equation}
	and $\var\{\psi^{\textup{tr}}_{\text{\sc cdr}}(O)\}$ is minimized uniquely by setting $p$ equal to $p_{\textup{opt}}$ defined by \eqref{opt.cdr}.
\end{thm}

Note that the optimal CDR design does not depend on $F^*$ whereas the optimal CIR design does. If $F^*=F$, then $r(W)\equiv1$ and $\pi_{\textup{opt}}^*=\pi_{\textup{opt}}$; otherwise, $\pi_{\textup{opt}}^*$ may differ from $\pi_{\textup{opt}}$.

\subsection{The Case of Generalization}\label{gen}

In this case, $F^*$ is described by observed covariate values (denoted by $W_j^*$, $j=1,\dots,N$) in a target cohort, a random sample from the target population that contains the trial cohort as a subset. Note that the trial cohort need not be a random sub-sample of the target cohort. For a generic patient in the target cohort, let $Z$ be an indicator for trial participation ($Z=1$ if in the trial; 0 otherwise), and let $X=(W,A,Y)=(W^*,A,Y)$ (if $Z=1$) or $W^*$ (if $Z=0$). In the present notation, $W^*\sim F^*$ and $(W^*|Z=1)\sim W\sim F$. The available data, $\{(W_i,A_i,Y_i),i=1,\dots,n\}\cup\{W_j^*:Z_j=0\}$, can be considered independent copies of $O=(Z,X)$. 

Under CIR,  \citet{d19} derived the efficient influence function for estimating $\Delta^*$ as
$$
\psi^{\textup{gen}}_{\text{\sc cir}}(O)=\delta(W^*)-\Delta^*
+\frac{ZA\{Y-m(1,W^*)\}}{e(W^*)\pi}
-\frac{Z(1-A)\{Y-m(0,W^*)\}}{e(W^*)(1-\pi)},
$$
where the superscript gen represents generalization and $e(W^*)=\pr(Z=1|W^*)$ is the propensity score for trial participation. Assumption \ref{eq.supp} implies that $\pr\{0<e(W^*)<1\}=1$. This result has been used to construct doubly robust, locally efficient estimators of $\Delta^*$ based on working parametric models for $(m,e)$ \citep{d19}.

\begin{lem}
In the case of generalization, the efficient influence function for estimating $\Delta^*$ under CDR is given by
$$
\psi^{\text{gen}}_{\text{\sc cdr}}(O)=\delta(\bmW^*)-\Delta^*
+\frac{ZA\{Y-m(1,\bmW^*)\}}{e(\bmW^*)p(\bmW^*)}
-\frac{Z(1-A)\{Y-m(0,\bmW^*)\}}{e(\bmW^*)\{1-p(\bmW^*)\}}.
$$
\end{lem}

Minimizing $\var\{\psi^{\text{gen}}_{\text{\sc cir}}(O)\}$ and $\var\{\psi^{\text{gen}}_{\text{\sc cdr}}(O)\}$ with respect to $\pi$ and $p$, respectively, yields the optimal CIR and CDR designs for generalization. For comparability (with the other cases), the optimal CIR design will be expressed using the general notation $r(\bmW)$, which in the present setting is inversely proportional to $e(\bmW)$: $r(\bmW)=\pr(Z=1)/e(\bmW)$.

\begin{thm}
Under Assumption \ref{eq.supp}, $\var\{\psi^{\text{gen}}_{\text{\sc cir}}(O)\}$ is minimized uniquely by setting $\pi$ equal to $\pi_{\text{opt}}^*$ defined by \eqref{opt.cir.tr}, and $\var\{\psi^{\text{gen}}_{\text{\sc cir}}(O)\}$ is minimized uniquely by setting $p$ equal to $p_{\text{opt}}$ defined by \eqref{opt.cdr}.
\end{thm}

Thus, despite the use of a different sampling mechanism, the optimal treatment allocations for generalization are identical to those for transportation.

\subsection{The Case of Post-Stratification}\label{ps}

In this case, $F^*$ is not described by individual patient data but by a known stratification with known weights. Let $\mathcal W$ be partitioned as $\cup_{k=1}^K\mathcal W_k$ with $\mathcal W_k\cap\mathcal W_l=\emptyset$ for $k\not=l$, and let $(\tau_1^*,\dots,\tau_K^*)$ be a set of known weights such that $\sum_{k=1}^K\tau_k^*=1$ and $\tau_k^*>0$, $k=1,\dots,K$. For example, the $\mathcal W_k$'s may be demographic subgroups based on age, gender and/or race, and the $\tau_k^*$'s may be obtained from a census or a disease registry much larger than the trial cohort \citep{cs10}. Based on this information, $F^*$ may be defined as follows:
$$
F^*(\mathcal B)=\sum_{k=1}^K\tau_k^*\pr(W\in\mathcal B|W\in\mathcal W_k)
=\sum_{k=1}^K\frac{\tau_k^*F(\mathcal B\cap\mathcal W_k)}{F(\mathcal W_k)}
$$
for any measurable subset $\mathcal B$ of $\mathcal W$. It follows that $\Delta^*=\int\delta(w)\/\text{d}F^*(w)=\sum_{k=1}^K\tau_k^*\Delta_k$, where $\Delta_k=\epn\{Y(1)-Y(0)|W\in\mathcal W_k\}$, $k=1,\dots,K$. In words, $\Delta^*$ is a weighted average of subgroup-specific ATE's in the trial population.

Because the $\tau_k^*$'s are known constants, the efficient influence function for estimating $\Delta^*$ is a weighted average of those for estimating the $\Delta_k$'s. Let $O=(W,A,Y)$, let $I(\cdot)$ be the indicator function, and let $\tau_k=\pr(W\in\mathcal W_k)$, $k=1,\dots,K$.

\begin{lem}
	In the case of post-stratification, the efficient influence function for estimating $\Delta^*$ is 
	$$
	\psi^{\textup{ps}}_{\text{\sc cir}}(O)=\sum_{k=1}^K\frac{I(W\in\mathcal W_k)\tau_k^*}{\tau_k}
	\left[\delta(W)-\Delta_k+\frac{A\{Y-m(1,W)\}}{\pi}-\frac{(1-A)\{Y-m(0,W)\}}{1-\pi}\right]
	$$
	under CIR, and
	$$
	\psi^{\textup{ps}}_{\text{\sc cdr}}(O)=\sum_{k=1}^K\frac{I(W\in\mathcal W_k)\tau_k^*}{\tau_k}
	\left[\delta(W)-\Delta_k+\frac{A\{Y-m(1,W)\}}{p(W)}-\frac{(1-A)\{Y-m(0,W)\}}{1-p(W)}\right]
	$$
	under CDR.
\end{lem}

Efficient estimation of $\Delta^*$ can be approached by first estimating each $\Delta_k$ efficiently. The latter can be achieved by applying an efficient method for ATE estimation to the subgroup of trial participants with $W_i\in\mathcal W_k$. General approaches to efficient estimation of the ATE under both CIR and CDR are available in the causal inference literature \citep[e.g.,][]{vr03,vr11,z23}. The next result provides the optimal CIR and CDR designs for post-stratification. In the present setting, we have $r(W)=\sum_{k=1}^KI(W\in\mathcal W_k)\tau_k^*/\tau_k$.

\begin{thm}
	Under Assumption \ref{eq.supp}, $\var\{\psi^{\textup{ps}}_{\text{\sc cir}}(O)\}$ is minimized uniquely by setting $\pi=\pi_{\textup{opt}}^*$ defined by \eqref{opt.cir.tr}, and $\var\{\psi^{\textup{ps}}_{\text{\sc cir}}(O)\}$ is minimized uniquely by setting $p=p_{\textup{opt}}$ defined by \eqref{opt.cdr}.
\end{thm}

In the three cases considered so far, the optimal CIR design depends on $F^*$ but not on the configuration of data or information describing $F^*$, and the optimal CDR design does not even depend on $F^*$.

\subsection{Extension to Other Effect Measures}\label{ext}

The results in Sections \ref{tr}--\ref{ps} can be extended to other effect measures of the form $\Delta^*(g)=g(\mu_1^*)-g(\mu_0^*)$, where $\mu_a^*=\int m(a,w)\/\text{d}F^*(w)$, $a=0,1$, and $g$ is a specified differentiable and increasing function. If $g$ is the identity function, then $\Delta^*(g)=\Delta^*$. Other examples of $g$ include the log and logit functions. Under a suitable mean exchangeability assumption \citep[e.g.,][]{z16,d19,c24}, $\mu_a^*$ represents the mean potential outcome for treatment $a\in\{0,1\}$ in the target population, and $\Delta^*(g)$ is a contrast of $\mu_1^*$ with $\mu_0^*$. We assume that Assumption \ref{eq.supp} holds and write $g'$ for the derivative function of $g$. In each of the three cases considered (transportation, generalization and post-stratification), the optimal CIR design for estimating $\Delta^*(g)$ is 
$$
\pi_{\textup{opt}}^*(g)=\frac{g'(\mu_1^*)[\epn\{r(W)^2v_1(W)\}]^{1/2}}{g'(\mu_1^*)[\epn\{r(W)^2v_1(W)\}]^{1/2}+g'(\mu_0^*)[\epn\{r(W)^2v_0(W)\}]^{1/2}},
$$
and the optimal CDR design is $p_{\textup{opt}}^*(W;g)=\{g'(\mu_1^*)v_1(W)^{1/2}\}/\{g'(\mu_1^*)v_1(W)^{1/2}+g'(\mu_0^*)v_0(W)^{1/2}\}$. These results follow from simple modifications of the proofs in Supplementary Materials, and the details are omitted. If $g$ is nonlinear, the optimal CDR design for estimating $\Delta^*(g)$ generally depends on $F^*$ through $(\mu_0^*,\mu_1^*)$.

\section{Numerical Results}\label{nr}

\subsection{Simulation}\label{sim}

Imagine we are designing a randomized trial in a specific patient population with a continuous outcome and two important baseline covariates. Specifically, $W=(W_1,W_2)'$ with $W_1\sim N(0,0.75^2;-2,2)$ and $W_2\sim B(0.2)$, independent of each other. Here and in the sequel, $N(\mu,\sigma^2;l,u)\sim(X|l\le X\le u)$ with $X\sim N(\mu,\sigma^2)$ and $B(q)$ denotes the Bernoulli distribution with success probability $q$. Given $W$, the potential outcomes are distributed as follows: 
\begin{equation*}\begin{aligned}
		Y(0)&\sim N(W_1+W_2,\exp(-2+W_1+2W_2)),\\
		Y(1)&\sim N(1+W_2,\exp(1-W_1-2W_2)).
\end{aligned}\end{equation*}
The possible dependence of $Y(0)$ and $Y(1)$ given $W$ is irrelevant and left unspecified.

The immediate objective of the trial is to estimate the ATE in the trial population, $\Delta=1$. The trial data may also be used for transportation, generalization, or post-stratification. In the case of transportation, the target population is defined by $W_1^*\sim N(0.5,1;-2,2)$ and $W_2^*\sim B(0.5)$, independent of each other, and the estimand is $\Delta_{\textup{tr}}^*\approx0.621$. In the case of generalization, the target population is a 1:1 mixture of the trial population and the target population for transportation. To be precise, we may write $F_{\textup{gen}}^*=0.5F+0.5F_{\textup{tr}}^*$, where $F_{\textup{gen}}^*$ and $F_{\textup{tr}}^*$ are the target covariate distributions for generalization and transportation, respectively. Note that, under $F_{\textup{gen}}^*$, $W_1^*$ and $W_2^*$ are no longer independent of each other. The estimand in this case is $\Delta_{\textup{gen}}^*\approx0.810$. In the case of post-stratification, we partition $\mathcal{W}=[-2,2]\times\{0,1\}$ into four strata: $\mathcal{W}_1=[-2,0.5)\times\{0\}$, $\mathcal{W}_2=[0.5,2]\times\{0\}$, $\mathcal{W}_3=[-2,0.5)\times\{1\}$, and $\mathcal{W}_4=[0.5,2]\times\{1\}$. The associated weights are $(\tau_1^*,\ldots,\tau_4^*)= (0.1, 0.2, 0.3, 0.4)$, and the resulting estimand value is $\Delta_{\textup{ps}}^*\approx0.567$.

\begin{figure}[H]
	\centering
	\includegraphics[width=0.6\textwidth]{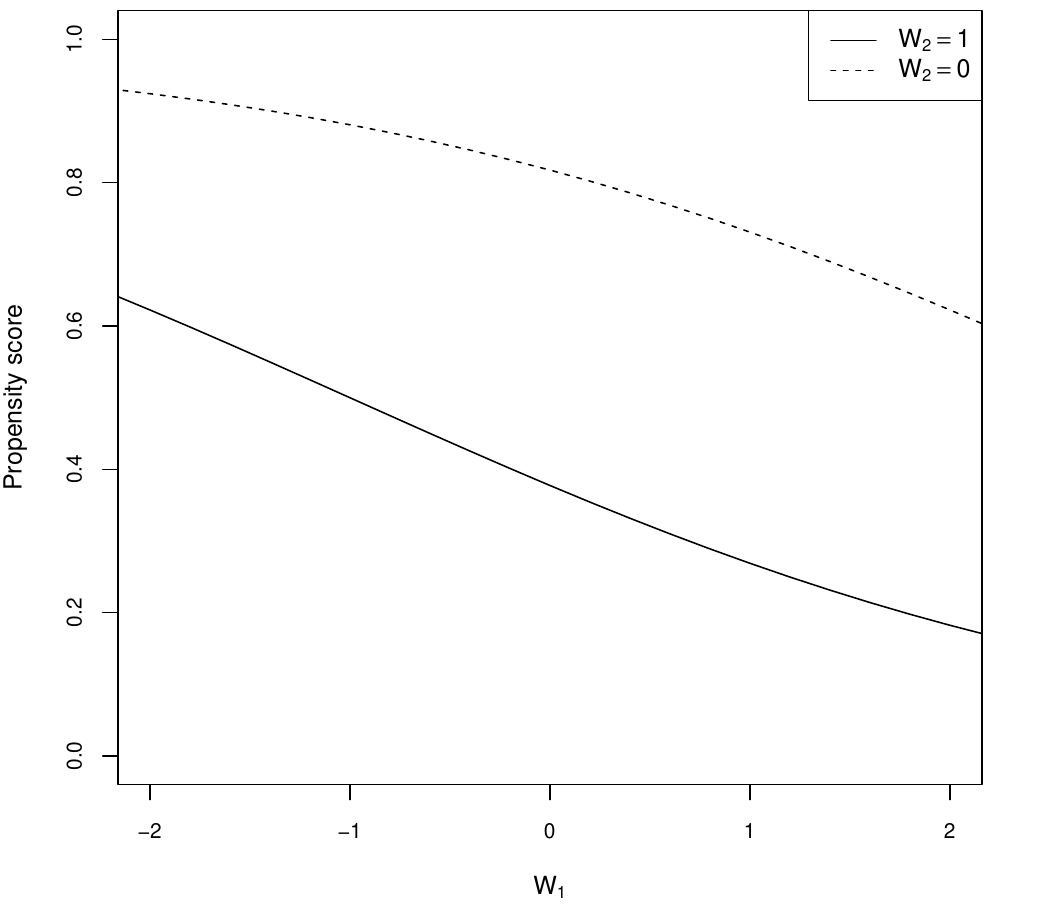} 
	\caption{The optimal CDR design in the simulation study.}
	\label{fig:opt.ps}
\end{figure}

As noted in Section \ref{ota}, the four estimands share the same optimal CDR design, which is depicted in Figure \ref{fig:opt.ps}. The four estimands do correspond to different optimal CIR designs, which are shown in Table \ref{sim.rst.re}. We also consider 1:1 CIR as a comparator. For each of the six designs, we set $n=250$ for the trial cohort and $n^*=250$ for transportation, and sample $N$ from a negative binomial distribution (with the goal of enrolling $n$ patients into the trial) in the case of generalization. Each design is simulated 5,000 times, and the resulting data are analyzed to estimate the four estimands using efficient estimators (described in Supplementary Materials).

As expected, all estimators are virtually unbiased under all designs (results not shown). Table \ref{sim.rst.re} compares the six designs in terms of efficiency with the 1:1 CIR design as the reference. The relative efficiency of an optimal design is calculated as the variance ratio of the 1:1 CIR design to the design in question. For each estimand, the corresponding optimal CIR design is indeed more efficient than the other CIR designs. The relative efficiency results of the optimal CIR designs vary widely across the four estimands. In contrast, the optimal CDR design attains the highest level of efficiency for all four estimands, as predicted by theory.

\begin{table}[H]
		\caption{Relative efficiency results in the simulation study, with 1:1 allocation as the reference.}\label{sim.rst.re}
		\newcolumntype{d}{D{.}{.}{3}}
		\newcolumntype{e}{D{.}{.}{1}}
		\begin{center}
			\begin{tabular}{ccccc}
				\hline
				\hline
Allocation &$\Delta$ &$\Delta^*_{\textup{tr}}$ &$\Delta^*_{\textup{gen}}$&$\Delta^*_{\textup{ps}}$\\
\hline
$\pi=0.5$             &1      &1       &1        &1\\
$\pi_{\textup{opt}}\approx0.730$&1.160  &0.645  &0.797   &0.636\\
$\pi_{\textup{opt}}^{\textup{tr}}\approx0.291$&0.649  &1.086  &0.891   &1.179\\
$\pi_{\textup{opt}}^{\textup{gen}}\approx0.460$&0.910  &1.036  &1.029   &1.100\\
$\pi_{\textup{opt}}^{\textup{ps}}\approx0.260$&0.564  &1.050  &0.834   &1.182\\
$p_{\text{opt}}(W)$&1.355  &1.241  &1.389   &1.240\\
	\hline
			\end{tabular}
		\end{center}
	{\footnotesize  $\Delta$: ATE in the trial population; $\Delta^*_{\textup{tr}}$: estimand for transportation; $\Delta^*_{\textup{gen}}$: estimand for generalization; $\Delta^*_{\textup{ps}}$: estimand for post-stratification.}
\end{table}

\subsection{Example}\label{ex}

We now illustrate our findings by applying them retrospectively to a completed trial. The BENCHMRK trial is a randomized placebo-controlled study of raltegravir, an HIV-1 integrase inhibitor, for treating human immunodeficiency virus type 1 (HIV-1) in treatment-experienced patients \citep{s08,c08}. This trial randomized 699 HIV-1 patients in a 2:1 ratio to receive raltegraviror or placebo, each combined with optimized background therapy. The outcome measure of interest to us is the virologic response rate at week 48 of treatment, defined as the proportion of patients with viral load below 50 copies/ml, with discontinuation counted as failure. The observed virologic response rate is 60.4\% in the raltegraviror group and 30.8\% in the placebo group, for a difference of 29.6\% (95\% confidence interval: 22.2--37.0\%), which is significantly greater than 0 (one-sided p value $< 0.0001$).

We first calculate optimal treatment allocations for estimating the ATE of raltegravir in the BENCHMRK trial population. The covariate vector $W$ consists of race (black or not), age, body mass index, log-transformed viral load and CD4 cell count, as well as genotypic and phenotypic sensitivity scores (defined as the number of concurrently used antiretroviral drugs to which a subject's HIV is fully susceptible, based on genotypic or phenotypic resistance testing). In this calculation, we assume that $W$ is distributed as observed in the BENCHMRK trial (excluding 37 patients with missing covariate data) and that, given $W$, the potential outcomes follow a logistic regression model:
\begin{equation}\label{logitY}
	\logit[\pr\{Y(a)=1\mid W\}]=\beta_0+\beta_1a+\beta_2^TW+\beta_3^T(aW),\quad a=0,1.
\end{equation}
The parameter values in this model are equated to their estimates from fitting the observed-data model
$\logit\{\pr(Y=1 \mid A, W)\}=\beta_0+\beta_1A+\beta_2^TW+\beta_3^T(AW),$
to the actual trial data. This calculation yields $\pi_{\textup{opt}}\approx0.540<2/3$.

Next, we consider transporting the BENCHMRK trial results to another population. The target population is defined by the SAILING trial, a randomized non-inferiority trial comparing dolutegravir, a newer HIV-1 integrase inhibitor, with raltegravir for treating HIV-1 in treatment-experienced patients \citep{c13}. Such a transportation analysis is useful for making an indirect comparison of dolutegravir with placebo \citep{z16}. This calculation assumes the same outcome model \eqref{logitY} as well as a log-linear model for the density ratio: $\log\{r(w)\}=\alpha_0+\alpha_1^Tw$, with parameter values estimated from a logistic regression analysis of trial membership conditional on covariates \citep{z16}. The optimal value of $\pi$ is found to be $\pi_{\textup{opt}}^{\textup{tr}}\approx0.481$.

A transportation analysis typically requires access to individual patient data in both trials. Without access to individual patient data from SAILING, one could use the published SAILING results \citep{c13}, together with individual patient data from BENCHMRK, to perform a post-stratification analysis that partially adjusts for population differences. For illustration, we consider post-stratification on the phenotypic sensitivity score, a known effect modifier \citep[Figure 1]{c08} which appears differentially distributed between the two trials \citep[Table 3]{z16}. The weights of the strata are taken to be the observed proportions in SAILING. Under the same assumptions for calculating $\pi_{\textup{opt}}$, the optimal value of $\pi$ for post-stratification is found to be $\pi_{\textup{opt}}^{\textup{ps}}\approx0.528$.

Table \ref{ex.rst.re} reports an efficiency comparison of the optimal CDR design, the three optimal CIR designs described above, and the original 2:1 CIR design. For a given estimand, the relative efficiency of each optimal design is calculated theoretically as its information bound divided by the information bound for the original design. The results in Table \ref{ex.rst.re} are consistent with theoretical predictions as well as the simulation results.

\begin{table}[H]
		\caption{Relative efficiency results in the HIV example, with the original 2:1 allocation as the reference.}\label{ex.rst.re}
		\newcolumntype{d}{D{.}{.}{3}}
		\newcolumntype{e}{D{.}{.}{1}}
		\begin{center}
			\begin{tabular}{cccc}
				\hline
				\hline
Allocation &$\Delta$ &$\Delta^*_{\textup{tr}}$ &$\Delta^*_{\textup{ps}}$\\
\hline
$\pi=2/3$             &1       &1        &1\\
$\pi_{\textup{opt}}\approx0.540$&1.071  &1.139  &1.084\\
$\pi_{\textup{opt}}^{\textup{tr}}\approx0.481$&1.057  &1.154  &1.075\\
$\pi_{\textup{opt}}^{\textup{ps}}\approx0.528$&1.071  &1.144  &1.085\\
$p_{\text{opt}}(W)$&1.101  &1.172  &1.109\\
	\hline
			\end{tabular}
		\end{center}
	{\footnotesize  $\Delta$: ATE in the trial population; $\Delta^*_{\textup{tr}}$: estimand for transportation; $\Delta^*_{\textup{ps}}$: estimand for post-stratification.}
\end{table}

\section{Concluding Remarks}\label{cr}

It is well known that, for any given effect measure, the optimal CDR design is generally more efficient than the optimal CIR design because CDR includes CIR as a special case \citep{z23}. The results in the present article indicate that the optimal CDR design may also be less sensitive to the target population for treatment effect estimation. Indeed, for ATE estimation, there is a single CDR design that is optimal for any target population. For the nonlinear effect measures considered in Section \ref{ext}, the optimal CDR design depends on the target population through $(\mu_0^*,\mu_1^*)$, but the dependence is relatively simple and does not involve the (typically) unknown density ratio $r$. These observations provide additional support for the use of optimal CDR designs in clinical trials.

\section*{Acknowledgements}

The research of Wei Zhang was supported by the National Key R\&D Program of China [grant number 2022YFA1004800]. The research of Aiyi Liu was supported by the intramural research program of the \textit{Eunice Kennedy Shriver} National Institute of Child Health and Human Development.

\end{document}